\RequirePackage{rotating}
\documentclass[sigconf]{acmart}
\usepackage[htt]{hyphenat}
\usepackage{rotating}
\usepackage{graphicx}
\usepackage{caption}
\usepackage{subcaption}
\usepackage[htt]{hyphenat}

\AtBeginDocument{%
  \providecommand\BibTeX{{%
    \normalfont B\kern-0.5em{\scshape i\kern-0.25em b}\kern-0.8em\TeX}}}

\copyrightyear{2023}
\acmYear{2023}
\setcopyright{rightsretained}
\acmDOI{10.1145/3569951.3597583}

\acmConference[PEARC '23]{Practice and Experience in Advanced Research Computing}{July 23--27, 2023}{Portland, OR, USA}
\acmBooktitle{Practice and Experience in Advanced Research Computing (PEARC '23), July 23--27, 2023, Portland, OR, USA}
\acmISBN{978-1-4503-9985-2/23/07}
\begin{document}

\title{A Further Study of Linux Kernel Hugepages
on A64FX with FLASH, an Astrophysical
Simulation Code}

\author{Catherine Feldman}
\email{catherine.feldman@stonybrook.edu}
\author{Smeet Chheda}
\email{smeetdinesh.chheda@stonybrook.edu}
\author{Alan C.\ Calder}
\email{alan.calder@stonybrook.edu}
\author{Eva Siegmann} 
\email{eva.siegmann@stonybrook.edu}
\author{John Dey} 
\email{john.dey@stonybrook.edu}
\author{Tony Curtis}
\email{anthony.curtis@stonybrook.edu}
\author{Robert J.\ Harrison}
\email{robert.harrison@stonybrook.edu}

\affiliation{%
  \institution{Institute for Advanced Computational Science}
  \streetaddress{Stony Brook University}
  \city{Stony Brook}
  \state{New York}
  \country{USA}
  \postcode{11794-5250}
}

\renewcommand{\shortauthors}{Feldman, et al.}

\begin{abstract}
We present an expanded study of the performance of FLASH when using Linux Kernel Hugepages
on Ookami, an HPE Apollo 80 A64FX platform. FLASH is a multi-scale, 
multi-physics simulation code written principally
in modern Fortran and makes use of the PARAMESH library to manage a block-structured
adaptive mesh. Our initial study used only the Fujitsu compiler to utilize standard
hugepages (hp), but further investigation allowed us to utilize hp for multiple
compilers by linking to the Fujitsu library \texttt{libmpg} and 
transparent hugepages (thp) by enabling it at the node level. By comparing the results of hardware counters and in-code timers, we found that hp and thp do not significantly impact the runtime performance of FLASH. Interestingly, there is a significant reduction in the TLB misses, differences in cache and memory access counters, and strange behavior is observed when using thp.

\end{abstract}





%\received{20 April 2023}
%\received[revised]{XX Month 2023}
%\received[accepted]{XX Month 2023}

\maketitle

\section{Introduction}

\subsection{Ookami and A64FX}

The A64FX processor expects to provide high 
performance and reliability for memory-intensive applications while
maintaining a good performance to power ratio. The appeal of A64FX, currently the backbone of the Fugaku supercomputer, is that it eliminates the need to port to accelerators such as GPUs to improve performance.
Ookami is an open-access resource featuring Fujitsu A64FX processors 
provided under the US NSF's 
ACCESS program and managed jointly by
Stony Brook University and  the University at Buffalo.
Ookami is an HPE/Cray Apollo80 system with 176 A64FX Fujitsu compute nodes, each with 32GB high-bandwidth
memory (HBM) and a 512GB SSD.
Ookami's FX700 series A64FX processors consist of four core memory
groups each with 12 cores, resulting in a total of 48 cores, 64KB L1 cache per core, and 8MB L2 cache shared
between the cores and runs at 1.8 GHz. The nodes have 32 GB of high-bandwidth memory, where 5 GB are reserved for the OS, leaving 27 GB for the user. These processors use the ARMv8.2--A
Scalable Vector Extension (SVE) SIMD instruction set with a 512 bit vector
implementation, allowing for vector lengths anywhere from 128--2048 bits
and enabling vector length agnostic programming \cite{pearc_experiences_2021}.

\subsection{Thermonuclear Supernovae with FLASH}
Our application is a bright stellar explosion known as
a thermonuclear (Type Ia) 
supernova (SN Ia), which we model using 
FLASH, a
software instrument for addressing
multi-scale, multi-physics applications
~\cite{Fryxetal00}.
FLASH is written in modern Fortran, parallelized through MPI, and implements 
AMR (Adaptive Mesh Refinement) using the PARAMESH library.
Full-star hydrodynamics simulations such as these are memory and computationally intensive, making our application a good candidate to try on A64FX.
Early study of the performance of FLASH on Ookami may be found in~\cite{flashexperience2022}, and similar experiences
are reported in~\cite{bari_etal2021,domke2021}. The unoptimized
performance on A64FX did not compare well to that found on 
traditional X86 architectures \cite{pearc_experiences_2021}.

Profiling 
indicated that FLASH 
spent about half of its time in the hydrodynamics routines, 
and within those 20\% of the time was spent in routine for
the material equation of state (EOS) \cite{flashexperience2022}. 
We therefore settled on two test problems for further exploration: a 2-d SN Ia problem
(that exercises the material EOS) and, looking ahead to our 
science goal of 3-d SN Ia simulations,
a 3-d hydrodynamics simulation, the Sedov explosion problem.
We dubbed these two tests ``EOS'' and ``3-d Hydro'', and 
details of both the EOS and hydrodynamics modules 
may be found in the original FLASH paper~\cite{Fryxetal00}.

Our motivation for investigating huge memory pages was both the observed bountiful DTLB misses, and FLASH's memory stride.
PARAMESH manages a block-structured adaptive mesh, where each block is separated into smaller cells that each store requisite variables, such as density and temperature, consecutively in an array. Thus there is a stride in memory when gathering the same variable (i.e. density) from different cells, and a larger stride between blocks. 
%Also, FLASH uses double precision arithmetic.

\subsection{Previous Work with Hugepages}

Here, we explore both standard and transparent hugepages. Modern processors manage memory in blocks known as pages. Hugepage support was integrated into the Linux kernel in version 2.6. These pages are larger in size than regular pages, which in theory means 
there are fewer pages for the OS to manage as there is a finite amount of memory. Depending on the OS, hugepages come in different sizes. Managing these pages can be challenging and at times require changes to application code. To that extent, Transparent HugePages were implemented in the Linux kernel where the the "transparent" hugepages are an abstraction layer managed by the kernel, where the kernel is responsible for their creation, management and use~\cite{THP}. Transparent hugepages are by default disabled on Ookami.

Other studies that have tested the performance effects of using hugepages on A64FX include \cite{domke2021}, \cite{Langarita_2023}, and \cite{L2_TLB_miss}, and suggest certain environment variable settings for best results. \cite{Langarita_2023} explicitly shows that the greatest speedup gain from enabling hugepages is seen
for a latency-bound section of their simulation, but is only a 1.11 $\times$ speedup. \cite{L2_TLB_miss} found that an increase in L2 TLB misses caused
performance degradation when using normal 64 KiB pages, but
didn’t affect the performance when using 2 MiB hugepages.

% alan fix of overflow above by adding "using" in front of hugepages
This work extends our initial study of using 
hugepages with just the Fujitsu compiler, which demonstrated that hugepages did not provide a significant speedup \cite{flash_huge_2022}.
Our speculation was that TLB misses might not make much of a difference because the A64FX has hardware to ameliorate the cost 
of TLB misses by avoiding OS calls, or because the FLASH data access patterns do not trigger a performance penalty.

\section{Testing Use of Hugepages}
We ran the ``EOS'' and ``3-d Hydro'' test problems, as described above. 
The EOS test ran a $\sim$ 1 GB 2-d SN Ia simulation for 50 time steps 
and the 3-d Hydro test ran a $\sim$ 9 GB Sedov explosion simulation 
for 2 time steps. Both tests were run on 1 and 12 cores. We used the round robin distribution of processors for the runs on 
12 cores because FLASH Morton 
orders the blocks to be 
spatially located together. Filling one core memory
group first will put blocks together but round robin spreads them as much as possible. We ran each test 7 times, removed the highest and lowest run times, and averaged the results from the remaining 5.
To investigate the effects of hugepages, we used the Fujitsu hardware 
counters \cite{a64fxmicro} of the Performance Application Programming Interface (PAPI)~\cite{papi} to  monitor cycles, 
TLB misses, and memory access, and used FLASH's internal timers to obtain runtimes.
Tests consisted of running the PAPI-instrumented code without hugepages (no hp), with 2MB standard hugepages (hp), and with 2MB transparent hugepages (thp). To use (t)hp, we linked the GCC and ARM compilers to Fujitsu's \texttt{libmpg} library, and used compiler flags for the Fujitsu compiler.
A detailed description of the runtime environment, including library versions, compiler flags, linking to PAPI and Fujitsu's \texttt{libmpg} library, and how to enable/disable (t)hp can be found in Appendix \ref{appendix_env}.

%This means rank 0 is bound to core 0 (core 0 of CMG 1), rank 1 is bound to core 12 (core 0 of CMG 2), rank 3 is bound to core ...
% if we are running out of space, the rank <> core mapping can be zapped.
%RANK: 0  CPU\_SET:    0; NUMA: 0  Socket: 0\\
%RANK: 1  CPU\_SET:   12; NUMA: 1  Socket: 1\\
%RANK: 2  CPU\_SET:   24; NUMA: 2  Socket: 2\\
%RANK: 3  CPU\_SET:   36; NUMA: 3  Socket: 3\\
%RANK: 4  CPU\_SET:    1; NUMA: 0  Socket: 0\\
%RANK: 5  CPU\_SET:   13; NUMA: 1  Socket: 1\\
%RANK: 6  CPU\_SET:   25; NUMA: 2  Socket: 2\\
%RANK: 7  CPU\_SET:   37; NUMA: 3  Socket: 3\\
%RANK: 8  CPU\_SET:    2; NUMA: 0  Socket: 0\\
%RANK: 9  CPU\_SET:   14; NUMA: 1  Socket: 1\\
%RANK:10  CPU\_SET:   26; NUMA: 2  Socket: 2\\
%RANK:11  CPU\_SET:   38; NUMA: 3  Socket: 3\\
%
%To do this with MVAPICH:
%\texttt{export MV2\_SHOW\_CPU\_BINDING=1}
%\texttt{export MV2\_CPU\_BINDING\_POLICY=hybrid}
%\texttt{export MV2\_HYBRID\_BINDING\_POLICY=numa}
%\texttt{srun ./flash4}
%
%To do this with OpenMPI and Fujitsu compiler:
%\texttt{mpiexec --rank-by numa --map-by numa:pe=1 --report-bindings ./flash4}
%

\section{Results}
\begin{figure*}[ht!]
    \begin{subfigure}{.46\textwidth}
        \includegraphics[width=\textwidth]{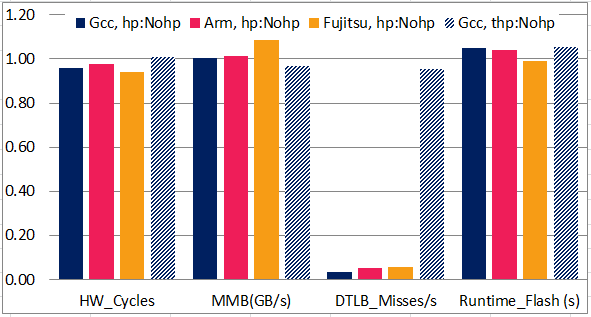}
        \caption{EOS test}
        \label{fig:EOS1p}
    \end{subfigure}%
    \begin{subfigure}{.54\textwidth}
        \includegraphics[width=\textwidth]{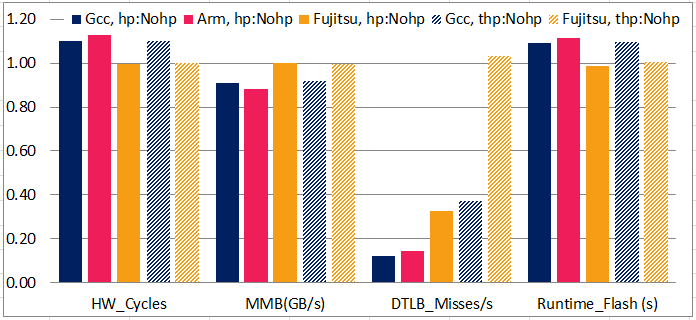}
        \caption{3-d hydro test}
        \label{fig:3DHydro1p}
    \end{subfigure}
    \caption{Ratios of runs with and without hugepages for each compiler for the (a) EOS test and (b) 3-d hydro test on 1 core}
\end{figure*}

First, we saw how the runtime, main memory bandwidth (MMB), and DTLB miss rate changed with huge page use. To do this, we used the following PAPI counters by setting \texttt{ PAPI\_EVENTS} to  \texttt{PERF\_COUNT\_HW\_CPU\_CYCLES,PERF\_COUNT\_HW\_CACHE\_MISSES,\-DTLB-L\-OAD-MISSES}.
%above alan fix of "\-" breaking LOAD. Uses hyphen package added above.
%,PERF\_COUNT\_HW\_STALLED\_CYC\-LES\_BACKEND,PERF\_COUNT\_HW\_STALLED\_CYCLES\-\_FRONTEND
The results from the 1 processor runs are shown for the EOS test in Figure \ref{fig:EOS1p}, and for the 3-d hydro in Figure \ref{fig:3DHydro1p} -- the 12 core runs exhibited similar patterns and are therefore not shown.

\begin{table*}[ht!]
\caption{Counters and derived rates for single core runs, for each test problem and two compilers. Values shown are ratios with : without standard hugepages enabled. Counter descriptions and rate calculations can be found in \cite{a64fxmicro}.}
\begin{tabular}{l|rr|rr}
 & {EOS} &  & {3-d Hydro} &  \\
Description & {GCC} & {Fujitsu} & {GCC} & {Fujitsu} \\
\hline
\hline
\texttt{DTLB-LOAD-MISSES} & 0.03 & 0.06 & 0.11 & 0.31 \\
\texttt{L1D\_TLB\_REFILL} & 0.03 & 0.05 & 0.11 & 0.31 \\
\texttt{L2D\_TLB\_REFILL} & 0.0002 & 0.01 & 0.03 & 0.03 \\
\texttt{L1I\_TLB\_REFILL} & 0.71 & 1.01 & 0.04 & 0.65 \\
\texttt{L2I\_TLB\_REFILL} & 1.00 & 0.99 & 0.59 & 0.16 \\
\hline
\texttt{L1D\_CACHE\_REFILL} & 0.96 & 0.99 & 1.00 & 1.00 \\
\texttt{L2D\_CACHE\_REFILL} & 1.08 & 1.06 & 0.96 & 1.03 \\
\texttt{LD\_COMP\_WAIT} & 0.71 & 0.78 & 1.17 & 0.99 \\
\texttt{LD\_COMP\_WAIT\_L1\_MISS} & 0.82 &  0.78 & 0.94 & 1.00 \\
\texttt{LD\_COMP\_WAIT\_L2\_MISS} & 0.90 & 0.96 & 0.97 & 0.98 \\
\hline
Average latency of L1D cache miss processing & 1.03 & 1.03 & 1.00 & 1.00\\
Average latency of L2 cache miss processing & 2.53 & 1.00 & 1.03 & 0.96 \\
Bidirectional effective bandwidth between L1D cache and L2 cache & 1.01 & 1.07 & 0.91 & 1.00\\
Bidirectional effective bandwidth between L2 cache and memory & 1.10 & 1.11 & 0.87 & 1.04 \\
\hline
\end{tabular}
 \label{tab:counters_hp}
\end{table*}

The figures show the ratios of runs with and without (t)hp, e.g. values around 1 indicate no changes, values < 1 indicate a reduction by using (t)hp, and values > 1 an increase. It is important to note that only a portion of our code is instrumented with PAPI, namely the EOS calls for the EOS test, and the hydrodynamics calls for the 3-d hydro test. Therefore, these counters represent the behavior in that specific module, rather than the software as a whole, while the timers show the full runtime.
As expected and seen in our last study \cite{flash_huge_2022}, in both cases the hardware cycles, MMB, and overall runtime are about the same when using hp, thp, or no hp. 
However, using hp drastically decreases the DTLB miss rate, while using thp does not have as much of an effect.

Using thp proved to be an interesting struggle. Thp would not enable in our 1 core runs with the Fujitsu compiler for the EOS test, and is therefore not shown in Figure \ref{fig:EOS1p}. We finally saw thp usage by mapping the process to NUMA node 1 instead of NUMA node 0. When running the 3-d hydro application compiled with GCC on 12 cores, the node would reset in the middle of execution when thp was enabled. These difficulties using thp will be investigated in the future.

We also observed the change in selected hardware counters and their derived rates when enabling hp. We found that most of these counters varied by only around 1\%, so we report ratios of counters from a single run rather than an average as before. A64FX has 6 hardware counters, so these results were collected across multiple runs. For ease of interpretation, we ran these exploratory tests on 1 core only. The ratios of hp : no hp for the most relevant values are shown in Table \ref{tab:counters_hp}, and full tables showing all measured counters and rates can be found in Appendix \ref{appendix_counters}. 
As before, values < 1 indicate a reduction by using hp, and values > 1 an increase.

As expected, the TLB-related counters showed the biggest change. Although the L2-DTLB showed the greatest improvement when hp was enabled, 99\% of the total DTLB misses resulted in an L1-DTLB miss, and only < 1\% resulted in a L2-DTLB miss. The instruction TLBs were less affected. GCC typically exhibited a greater decrease in TLB refills than Fujitsu. The runtime, number of L1D and L2D cache misses, and the bandwidth were relatively unaffected by hp use. For the EOS test, the number of cycles spent waiting for memory access completion (\texttt{LD\_COMP\_WAIT}) is smaller when hp is enabled, but for the GCC compiler, the latency of L2 cache miss processing is higher. For the 3-d Hydro test with GCC, enabling hp slightly increased the total number of cpu cycles as well as (\texttt{LD\_COMP\_WAIT}). Overall, enabling hp has the overwhelming effect of reducing TLB misses, but not much else. The Fujitsu compiler seems to have less prominent changes in its counters than GCC.

\begin{figure*}[ht!]
    \centering
    \includegraphics[width=\textwidth]{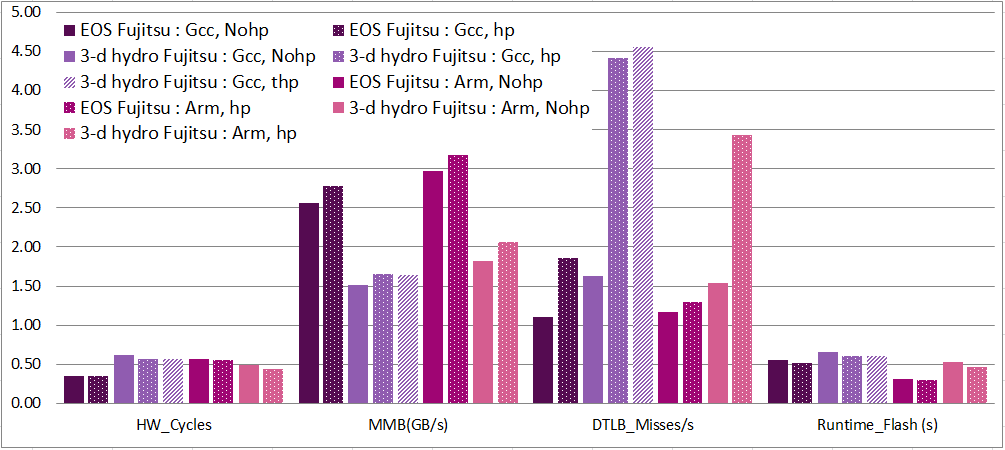}
    \caption{Ratios between the Fujitsu and other compilers (GCC and ARM), for each application and type of huge page.}
    \label{fig:compilers}
\end{figure*}

\begin{table*}[]
\caption{Counters and derived rates for single core runs, for each test problem with either standard hp or no hp enabled. Values shown are ratios for Fujitsu : GCC compiler. Counter descriptions and rate calculations can be found in \cite{a64fxmicro}.}
\begin{tabular}{lrrrr}
 & {EOS} &  & {3-d Hydro} &  \\
Description & {Hp} & {No hp} & {Hp} & {No hp} \\
\hline
\hline
\texttt{DTLB-LOAD-MISSES} & 0.66 & 0.39 & 2.20 & 0.82\\
\texttt{L1D\_TLB\_REFILL} & 0.55 & 0.39 & 2.52 & 0.86\\
\texttt{L2D\_TLB\_REFILL} & 0.77 & 0.02 & 0.93 & 1.02\\
\texttt{L1I\_TLB\_REFILL} & 0.70 & 0.49 & 0.63 & 0.04\\
\texttt{L2I\_TLB\_REFILL} & 1.00 & 1.01 & 0.64 & 2.33\\
\hline
\texttt{L1D\_CACHE\_REFILL} & 0.94 & 0.90 & 0.91 & 0.92 \\
\texttt{L2D\_CACHE\_REFILL} & 1.04 & 1.06 & 1.08 & 1.00\\
\texttt{LD\_COMP\_WAIT} & 0.50 & 0.46 & 0.66 & 0.78\\
\texttt{LD\_COMP\_WAIT\_L1\_MISS} & 0.56 & 0.58 & 2.58 & 2.43 \\
\texttt{LD\_COMP\_WAIT\_L2\_MISS} & 0.82 & 0.77 & 2.11 & 2.09\\
\hline
Average latency of L1D cache miss processing & 0.90 & 0.90 & 1.04 & 1.03\\
Average latency of L2 cache miss processing & 0.25 & 0.64 & 0.89 & 0.94 \\
Bidirectional effective bandwidth between L1D cache and L2 cache & 2.76 & 2.59 & 1.63 & 1.49\\
Bidirectional effective bandwidth between L2 cache and memory & 2.88 & 2.85 & 1.91 & 1.61 \\
\end{tabular}
\label{tab:counters_compiler}
\end{table*}

We also compared the single core results between compilers, namely to the Fujitsu compiler, which by far produced the fastest runtime. Figure \ref{fig:compilers} shows the ratio between the Fujitsu and other compilers (purple for GCC, pink for ARM) for each test problem (darker colors for EOS) and type of hugepage (solid for no hp, dotted for hp, and striped for thp), using the same dataset as that from Figures \ref{fig:EOS1p} and \ref{fig:3DHydro1p}. Here, values < 1 indicate a reduction due to use of the Fujitsu compiler, and values > 1 indicate an increase. Regardless of hugepage use, the Fujitsu compiler was nearly twice as fast as the others, and nearly four times as fast as ARM for the EOS test. The Fujitsu compiler also executes about half of the hardware cycles. For the EOS test, the Fujitsu compiler has a 2.5-3$\times$ greater MMB than the others; this is about 1.5-2$\times$ for 3-d Hydro. This is true even though the Fujitsu compiler exhibits a higher DTLB miss rate, which interestingly increases with huge page use. This rate increase says nothing about the relative TLB misses between the compilers, however, so for a better comparison we look at the ratios between the raw counter values and derived rates.

Table \ref{tab:counters_compiler} shows the ratio between the Fujitsu and GCC compilers of a subset of counters, for each test problem with hp and no hp enabled. We chose to compare only these two compilers since the ARM compiler is too slow to be a viable choice for production runs, and we only look at no hp and hp runs because thp did not even achieve the goal of reducing TLB misses. 
Again, values < 1 indicate a reduction due to use of the Fujitsu compiler, and values > 1 indicate an increase. 
The data used is the same as that used to create Table \ref{tab:counters_hp}, and full tables showing all measured counters and rates can be found in Appendix \ref{appendix_counters}.

Although the Fujitsu compiler has a much higher TLB miss rate than the GCC compiler in most cases, it has lower total TLB misses. The Fujitsu compiler also has a higher (1.6-2.9 $\times$) memory bandwidth and lower latency. It has the same number of cache misses, but spends less total cycles waiting for memory access than the GCC compiler.

\section{Summary and Conclusions} \label{sec:conclusion}
We found that for all compilers and both test problems, the use of both standard and transparent huge pages 
did not significantly affect the performance of FLASH, despite a drastic decrease in TLB misses. This suggests that TLB misses indeed do not have an impact on the performance. This may be due to the A64FX's translation table cache (TTC), which decreases the latency of virtual to physical address translation \cite{a64fxmicro}. Higher cache miss rates when using the Fujitsu compiler are offset by higher memory bandwidth and lower latency, which results in a shorter runtime. 

The Fujitsu compiler demonstrates 2-4 times better 
performance than the GCC and ARM compilers.
Although the Fujitsu compiler uses only half the total cycles of the GCC compiler, both compilers have the same number of cache misses. Since the bandwidth is $\sim$ 2 $\times$ larger for Fujitsu, this means that less time is spent waiting for memory access completion (ie in \texttt{LD\_COMP\_WAIT}), thereby shortening the runtime. However, only $\sim$ 20 \% - 40 \% of the total cycles are spent in \texttt{LD\_COMP\_WAIT}, so a higher bandwidth can't completely account for the faster runtime.
A contributing factor could be that Fujitsu may have better optimizations that take advantage of the A64FX hardware. This includes the use of SVE -- the Fujitsu executable uses the SVE registers 21 $\times$ more than GCC. The reason why Fujitsu produces the fastest executable, and what the performance bottlenecks are, will be explored in detail in future work.

\begin{comment}
{\color{teal}From the manual:

5. The LD flow performs address translation with L1D-TLB and in concurrently reads a tag and data
from L1D cache in 0th/1st-pipeline.. If the LD flow hits L1-DTLB and the L1D cache, it reads data
along the predefined pipeline stages and writes data in a register. The execution of the LD flow is
completed.

6. If the LD flow does not hit L1-DTLB in step 5, it searches L2-DTLB and then the translation table
to obtain virtual address translation information. If an L1D cache miss occurs, data fill is
performed by obtaining data from a lower cache level. In such cases, the MI flow and MO flow is
executed.

7. After the L1-DTLB or L1D cache miss is resolved, the LD flow is submitted from the RFP to
0th/1st-pipeline again and executed
}
\end{comment}

\begin{acks}
Ookami is supported by the US NSF grant \#1927880, and 
this research was supported in part by the US DOE 
under grant DE-FG02-87ER40317. FLASH was developed in part by the US
DOE NSA-ASC and OSC-ASCR-supported Flash Center for Computational 
Science at the University of Chicago.
The authors gratefully acknowledge the generous support 
of the Ookami community. The authors also
thank Jens Domke at RIKEN for very helpful suggestions.
\end{acks}

\bibliographystyle{ACM-Reference-Format}
\bibliography{bibliography}

\appendix
\section{Environment}
\label{appendix_env}
\subsection{Libraries and compiler flags}
We performed this study using FLASH version 4.6.2, including additional modules for our SN Ia application. To enable thp, we were conferred a dedicated node of Ookami
running 
Rocky Linux 8.4 with kernel 4.18.0-305.25.1.el8\_4.aarch64. To provide a more equal comparison, all runs used serial hdf5/1.10.1 and the same PAPI library. For this study, we were unable to get the Cray compiler to use the same
HDF5 library (necessary) as the other compilers, and therefore it is not used here, although our earlier study demonstrates that the Cray and GCC compilers give similar performance for FLASH. The performance of three compilers - GCC, Fujitsu, and ARM - were compared, and the compiler options for each are listed in Table \ref{tab:flash_compilers}. 

\begin{table*}[ht!]
  \caption{Compiler flags and MPI implementations used for this study}
  \centering
  \begin{tabular}{l c c l}
    \hline \hline
    Compiler & Compiler Flags & Linker Flags & MPI Implementation\\
    \hline
    GCC 12.2.0 & \texttt{-O3 -mcpu=a64fx -mtune=a64fx} & & MVAPICH 2.3.7 \\
     & \texttt{-fdefault-real-8 -fdefault-double-8}& &\\ 
     & \texttt{-Wuninitialized -fallow-argument-mismatch} & &\\
     \hline
    ARM 21.0 & \texttt{-c -O3 -armpl -mcpu=a64fx -mtune=a64fx -r8} & \texttt{-lamath} & MVAPICH 2.3.7 \\
     \hline
    Fujitsu 4.5 & \texttt{-KSVE,A64FX,ARMV8\_3\_A -Az -Kfast} & & Fujitsu built-in 1.0.21.01\\
    &  \texttt{-CcdII4 -CcdRR8} & & (based on OpenMPI)\\
     \hline
  \end{tabular}
  \label{tab:flash_compilers}
\end{table*}

%working on the line breaking a la
%https://tex.stackexchange.com/questions/299/how-to-get-long-texttt-sections-to-break
%\texttt{C:\textbackslash doc\-uments and set\-tings}
% ----- reducing following step to a single line in the following paragraph (for space)
We linked to PAPI at compile-time with \texttt{-L /opt/cray/pe/papi\-/6.0.0.4/lib -lpapi} 
and then at run-time linked the executable to the proper library like so: \texttt{export LD\_LIBRARY\_PATH=/opt/\-cray/pe/papi/6.0.0.4/lib:\$\{LD\_LIBRARY\_PATH\} }.

\subsection{Enabling huge pages}
Paging policy for the static data area, stack/thread stack area, and reserved dynamic memory areas is defined by \texttt{XOS\_MMM\_L\_PAGING\_POLICY}.
We set the paging policy to demand for all three areas (\texttt{export XOS\_MMM\_L\_PAGING\_POLICY=demand:demand:demand}) to ensure that memory used is within the NUMA memory region as much as possible.   

To use (t)hp, we linked the GCC and ARM compilers to Fujitsu's \texttt{libmpg} library by adding \texttt{-Wl,-T/opt/FJSVxos/\-mmm/util/bss-2mb.lds -L/opt/FJSVxos/mmm/lib64 -lmpg} to the compile and link flags. For the Fujitsu compiler, we added \texttt{-Klargepage} and \texttt{-Knolargepage} to turn (t)hp on and off, respectively.

Switching between different pages is controlled by the \texttt{XOS\_MMM\_L\_HPAGE\_TYPE} environment variable, when using Fujitsu's \texttt{libmpg}. While the documentation mentions that acceptable values are \texttt{none} or \texttt{hugetlbfs}, ~\cite{fugaku_codesign_report} mentions another possible value, \texttt{thp} for the variable on Fugaku (A64FX FX1000). This is viable on the FX700 system as well. Therefore there are three values for this environment variable -- \texttt{none} (No hp), \texttt{hugetlbfs} (default, enables hp), and \texttt{thp} (enables thp).

The kernel
should invoke thp on its own 
when it processes a file greater than 2 Gb. 
Thp can be enabled or disabled by selecting \texttt{[always]} or \texttt{[never]}, respectively, in \texttt{/sys/kernel/mm/tranparent\_hugepage/enabled}. We monitored the use of hugepages by the machine by looking at specific system variables 
in \texttt{/proc/meminfo}: \texttt{HugePages\_Total} should be nonzero when hp is in use; \texttt{AnonHugePages} should be nonzero when thp is in use; and both variables should be zero when using no hp.

\section{Counter Data} 
\label{appendix_counters}
For completeness, the tables below report the raw counter values and derived rates for both the Fujitsu and GCC compilers, with standard hp and without hp, for both test problems (``EOS'' and ``3-d Hydro'') on 1p. This data was used to create Tables \ref{tab:counters_hp} and \ref{tab:counters_compiler}. Table \ref{tab:app_EOS_Counter} shows the raw counter values for the EOS test; Table \ref{tab:app_EOS_Rate} shows the derived rates for the EOS test; Table \ref{tab:app_3dHydro_Counter} shows the raw counter values for the 3-d Hydro test; and Table \ref{tab:app_3dHydro_Rate} shows the derived rates for the 3-d Hydro test. Counter descriptions and rate calculations can be found in the Fujitsu Microarchitecture Manual \cite{a64fxmicro}.

\begin{sidewaystable*}
\centering
\caption{Raw counter values and their ratios for the EOS test on 1 core. The first 3 columns show the raw counter values for the GCC compiler with and without hp, followed by the counter ratio. The next three columns show the same, but for the Fujitsu compiler. The last 2 columns show the ratio of counter values between Fujitsu : GCC compiler.}
\label{tab:app_EOS_Counter}
\begin{tabular}{lrrr|rrr|rr}
\textbf{Counter} & \textbf{GCC} & {\textbf{GCC}} & {\textbf{GCC}} & {\textbf{Fujitsu}} & {\textbf{Fujitsu}} & \textbf{Fujitsu} & \textbf{Fuj : GCC} & \textbf{Fuj : GCC} \\
& \textbf{hp} & \textbf{no hp} & {\textbf{hp : no hp}} & \textbf{hp} & {\textbf{no hp}} & \textbf{hp : no hp} & \textbf{hp} & \textbf{no hp} \\
\hline \hline
\texttt{CPU\_CYCLES} & 341210316099 & 354894128279 & 0.96 & 118393395910 & 126126802860 & 0.94 & 0.35 & 0.36 \\
\texttt{DTLB-LOAD-MISSES} & 136916841 & 4200311763 & 0.03 & 89784864 & 1624074206 & 0.06 & 0.66 & 0.39 \\
\texttt{L1D\_TLB\_REFILL} & 135140728 & 4186687491 & 0.03 & 74337421 & 1628705951 & 0.05 & 0.55 & 0.39 \\
\texttt{L2D\_TLB\_REFILL} & 212 & 966342 & 0.0002 & 163 & 21269 & 0.01 & 0.77 & 0.02 \\
\texttt{L1I\_TLB\_REFILL} & 74981652 & 105364002 & 0.71 & 52578326 & 52073237 & 1.01 & 0.70 & 0.49 \\
\texttt{L2I\_TLB\_REFILL} & 6206064 & 6207292 & 1.00 & 6203536 & 6259401 & 0.99 & 1.00 & 1.01 \\
\hline
\texttt{L1\_MISS\_WAIT} & 41995961928 & 42462092793 & 0.99 & 35374847991 & 34746226643 & 1.02 & 0.84 & 0.82 \\
\texttt{L1D\_CACHE\_REFILL} & 1211001903 & 1261895471 & 0.96 & 1132822066 & 1141540325 & 0.99 & 0.94 & 0.90 \\
\texttt{L1D\_CACHE\_REFILL\_HWPRF} & 19441488 & 16435557 & 1.18 & 107749070 & 103581864 & 1.04 & 5.54 & 6.30 \\
\texttt{L1D\_CACHE\_REFILL\_PRF} & 21247269 & 16573641 & 1.28 & 107955640 & 103723552 & 1.04 & 5.08 & 6.26 \\
\texttt{L1D\_CACHE\_REFILL\_DM} & 1173058042 & 1194338101 & 0.98 & 1012021771 & 1008685191 & 1.00 & 0.86 & 0.84 \\
\texttt{L1D\_CACHE\_WB} & 568115212 & 570331006 & 1.00 & 570460305 & 548102563 & 1.04 & 1.00 & 0.96 \\
\hline
\texttt{L2\_MISS\_WAIT} & 14156941 & 5198312 & 2.72 & 3749562 & 3539840 & 1.06 & 0.26 & 0.68 \\
\texttt{L2D\_CACHE\_REFILL} & 1649116 & 1533521 & 1.08 & 1721394 & 1629720 & 1.06 & 1.04 & 1.06 \\
\texttt{L2D\_CACHE\_REFILL\_HWPRF} & 702309 & 646787 & 1.09 & 897424 & 805954 & 1.11 & 1.28 & 1.25 \\
\texttt{L2D\_CACHE\_REFILL\_PRF} & 702309 & 646787 & 1.09 & 900301 & 808715 & 1.11 & 1.28 & 1.25 \\
\texttt{L2D\_CACHE\_REFILL\_DM} & 946807 & 886734 & 1.07 & 821093 & 821005 & 1.00 & 0.87 & 0.93 \\
\texttt{L2D\_CACHE\_WB} & 910868 & 883074 & 1.03 & 836069 & 814766 & 1.03 & 0.92 & 0.92 \\
\hline
\texttt{LD\_COMP\_WAIT} & 43456361267 & 61463918085 & 0.71 & 21856142950 & 28021892897 & 0.78 & 0.50 & 0.46 \\
\texttt{LD\_COMP\_WAIT\_EX} & 14295343427 & 12725601925 & 1.12 & 4395919156 & 4817739974 & 0.91 & 0.31 & 0.38 \\
\texttt{LD\_COMP\_WAIT\_PFP\_BUSY} & 0 & 0 & N/A & 49 & 0 & N/A & N/A & N/A \\
\texttt{LD\_COMP\_WAIT\_L1\_MISS} & 14764291144 & 18042715909 & 0.82 & 8197400785 & 10457746706 & 0.78 & 0.56 & 0.58 \\
\texttt{LD\_COMP\_WAIT\_L2\_MISS} & 85681191 & 94715479 & 0.90 & 70471471 & 73306995 & 0.96 & 0.82 & 0.77 \\
\texttt{EU\_COMP\_WAIT} & 192299195248 & 189702560873 & 1.01 & 62110089168 & 63986117005 & 0.97 & 0.32 & 0.34 \\
\texttt{BR\_COMP\_WAIT} & 912951041 & 795494616 & 1.15 & 306967691 & 298776080 & 1.03 & 0.34 & 0.38 \\
\texttt{BR\_MIS\_PRED} & 163423745 & 167518949 & 0.98 & 40681971 & 42617939 & 0.95 & 0.25 & 0.25 \\
\hline
\texttt{EFFECTIVE\_INST\_SPEC} & 247030572236 & 241626826143 & 1.02 & 81247006439 & 81247004885 & 1.00 & 0.33 & 0.34 \\
\texttt{LD\_SPEC} & 55921382684 & 54938703722 & 1.02 & 20359738305 & 20359737979 & 1.00 & 0.36 & 0.37 \\
\texttt{BASE\_LD\_REG\_SPEC} & 15673422673 & 14690743711 & 1.07 & 2004533696 & 2004533370 & 1.00 & 0.13 & 0.14 
\end{tabular}
\end{sidewaystable*}

\begin{sidewaystable*}
\centering
\caption{Derived rates and their ratios for the EOS test on 1 core. The first 3 columns show the rates for the GCC compiler with and without hp, followed by the rate ratio. The next three columns show the same, but for the Fujitsu compiler. The last 2 columns show the ratio of rates between Fujitsu : GCC compiler.}
\label{tab:app_EOS_Rate}
\begin{tabular}{lrrr|rrr|rr}
\textbf{Rate} & \textbf{GCC} & {\textbf{GCC}} & {\textbf{GCC}} & {\textbf{Fujitsu}} & {\textbf{Fujitsu}} & \textbf{Fujitsu} & \textbf{Fuj : GCC} & \textbf{Fuj : GCC} \\
& \textbf{hp} & \textbf{no hp} & {\textbf{hp : no hp}} & \textbf{hp} & {\textbf{no hp}} & \textbf{hp : no hp} & \textbf{hp} & \textbf{no hp} \\
\hline \hline
Cycles per instruction (CPI) & 1.38 & 1.47 & 0.94 & 1.46 & 1.55 & 0.94 & 1.05 & 1.06 \\
Branch misprediction rate & 0.0007 & 0.0007 & 0.95 & 0.001 & 0.001 & 0.95 & 0.76 & 0.76 \\
\hline
L1-ITLB miss rate & 0.0003 & 0.0004 & 0.70 & 0.0006 & 0.0006 & 1.01 & 2.13 & 1.47 \\
L1-DTLB miss rate & 0.0005 & 0.02 & 0.03 & 0.0009 & 0.02 & 0.05 & 1.67 & 1.16 \\
L2-ITLB miss rate & 2.51E-05 & 2.57E-05 & 0.98 & 7.64E-05 & 7.70E-05 & 0.99 & 3.04 & 3.00 \\
L2-DTLB miss rate & 8.58E-10 & 4.00E-06 & 0.0002 & 2.01E-09 & 2.62E-07 & 0.01 & 2.34 & 0.07 \\
\hline
L1D cache miss rate & 0.005 & 0.005 & 0.94 & 0.01 & 0.01 & 0.99 & 2.84 & 2.69 \\
... attributable to demand access & 0.005 & 0.005 & 0.96 & 0.01 & 0.01 & 1.00 & 2.62 & 2.51 \\
... attributable to prefetch access & 8.60E-05 & 6.86E-05 & 1.25 & 0.001 & 0.001 & 1.04 & 15.45 & 18.61 \\
... attributable to software prefetch access & 7.31E-06 & 5.71E-07 & 12.79 & 2.54E-06 & 1.74E-06 & 1.46 & 0.35 & 3.05 \\
L2 cache miss rate & 6.68E-06 & 6.35E-06 & 1.05 & 2.12E-05 & 2.01E-05 & 1.06 & 3.17 & 3.16 \\
... attributable to demand access & 3.83E-06 & 3.67E-06 & 1.04 & 1.01E-05 & 1.01E-05 & 1.00 & 2.64 & 2.75 \\
... attributable to prefetch access & 2.84E-06 & 2.68E-06 & 1.06 & 1.11E-05 & 9.95E-06 & 1.11 & 3.90 & 3.72 \\
... attributable to software prefetch access & 0.00 & 0.00 & N/A & 3.54E-08 & 3.40E-08 & 1.04 & N/A & N/A \\
\hline
Average latency of & & & & & & & \\
... L1D cache miss processing & 34.68 & 33.65 & 1.03 & 31.23 & 30.44 & 1.03 & 0.90 & 0.90 \\
... L2 cache miss processing & 8.58 & 3.39 & 2.53 & 2.18 & 2.17 & 1.00 & 0.25 & 0.64 \\
Average number of outstanding misses & & & & & & & \\
... in L1D cache miss processing & 0.12 & 0.12 & 1.03 & 0.30 & 0.28 & 1.08 & 2.43 & 2.30 \\
... in L2 cache miss processing & 4.15E-05 & 1.46E-05 & 2.83 & 3.17E-05 & 2.81E-05 & 1.13 & 0.76 & 1.92 \\
Bidirectional effective BW (GB/s) & & & & & & & & \\
... between L1D cache and L2 cache & 2.40 & 2.38 & 1.01 & 6.63 & 6.17 & 1.07 & 2.76 & 2.59 \\
... between L2 cache and memory & 0.003 & 0.003 & 1.10 & 0.010 & 0.009 & 1.11 & 2.88 & 2.85 \\
\hline
\texttt{LD\_COMP\_WAIT/CPU\_CYCLES} & 0.13 & 0.17 & 0.74 & 0.18 & 0.22 & 0.83 & 1.45 & 1.28 \\
\texttt{LD\_COMP\_WAIT\_L1\_MISS/CPU\_CYCLES} & 0.04 & 0.05 & 0.85 & 0.07 & 0.08 & 0.84 & 1.60 & 1.63 \\
\texttt{LD\_COMP\_WAIT\_L2\_MISS/CPU\_CYCLES} & 0.0003 & 0.0003 & 0.94 & 0.0006 & 0.0006 & 1.02 & 2.37 & 2.18 \\
\texttt{LD\_COMP\_WAIT\_EX/CPU\_CYCLES} & 0.04 & 0.04 & 1.17 & 0.04 & 0.04 & 0.97 & 0.89 & 1.07 \\
\texttt{LD\_COMP\_WAIT\_PFP\_BUSY/CPU\_CYCLES} & 0.0 & 0.0 & N/A & 4.14E-10 & 0.00 & N/A & N/A & N/A
\end{tabular}
\end{sidewaystable*}

\begin{sidewaystable*}[]
\centering
\caption{Raw counter values and their ratios for the 3-d Hydro test on 1 core. The first 3 columns show the raw counter values for the GCC compiler with and without hp, followed by the counter ratio. The next three columns show the same, but for the Fujitsu compiler. The last 2 columns show the ratio of counter values between Fujitsu : GCC compiler.}
\label{tab:app_3dHydro_Counter}
\begin{tabular}{lrrr|rrr|rr}
\textbf{Counter} & \textbf{GCC} & {\textbf{GCC}} & {\textbf{GCC}} & {\textbf{Fujitsu}} & {\textbf{Fujitsu}} & \textbf{Fujitsu} & \textbf{Fuj : GCC} & \textbf{Fuj : GCC} \\
& \textbf{hp} & \textbf{no hp} & {\textbf{hp : no hp}} & \textbf{hp} & {\textbf{no hp}} & \textbf{hp : no hp} & \textbf{hp} & \textbf{no hp} \\
\hline \hline
\texttt{CPU\_CYCLES} & 2143932684764 & 1954934065987 & 1.10 & 1196335386888 & 1200402908968 & 1.00 & 0.56 & 0.61 \\
\texttt{DTLB-LOAD-MISSES} & 226322429 & 1979316404 & 0.11 & 498078094 & 1615854112 & 0.31 & 2.20 & 0.82 \\
\texttt{L1D\_TLB\_REFILL} & 200617382 & 1889017976 & 0.11 & 505910868 & 1623370615 & 0.31 & 2.52 & 0.86 \\
\texttt{L2D\_TLB\_REFILL} & 222571 & 6424782 & 0.03 & 207882 & 6556994 & 0.03 & 0.93 & 1.02 \\
\texttt{L1I\_TLB\_REFILL} & 5949138 & 141791086 & 0.04 & 3751271 & 5787082 & 0.65 & 0.63 & 0.04 \\
\texttt{L2I\_TLB\_REFILL} & 84205 & 141619 & 0.59 & 53741 & 330456 & 0.16 & 0.64 & 2.33 \\
\hline
\texttt{L1\_MISS\_WAIT} & 1174125092814 & 1175696041980 & 1.00 & 1112295369914 & 1114808777704 & 1.00 & 0.95 & 0.95 \\
\texttt{L1D\_CACHE\_REFILL} & 28765078635 & 28695255519 & 1.00 & 26312621913 & 26442346625 & 1.00 & 0.91 & 0.92 \\
\texttt{L1D\_CACHE\_REFILL\_HWPRF} & 7220868699 & 7128347854 & 1.01 & 4454501262 & 4526188824 & 0.98 & 0.62 & 0.63 \\
\texttt{L1D\_CACHE\_REFILL\_PRF} & 7746287019 & 7653780229 & 1.01 & 4590540067 & 4673132966 & 0.98 & 0.59 & 0.61 \\
\texttt{L1D\_CACHE\_REFILL\_DM} & 19022344796 & 19035542257 & 1.00 & 20327097179 & 20468323515 & 0.99 & 1.07 & 1.08 \\
\texttt{L1D\_CACHE\_WB} & 20130483048 & 20092819898 & 1.00 & 18212432757 & 18287841025 & 1.00 & 0.90 & 0.91 \\
\hline
\texttt{L2\_MISS\_WAIT} & 2310792458 & 2343447319 & 0.99 & 2211054362 & 2224081481 & 0.99 & 0.96 & 0.95 \\
\texttt{L2D\_CACHE\_REFILL} & 1827787623 & 1900351203 & 0.96 & 1975065788 & 1909551917 & 1.03 & 1.08 & 1.00 \\
\texttt{L2D\_CACHE\_REFILL\_HWPRF} & 1017222150 & 1024578925 & 0.99 & 963692147 & 897152822 & 1.07 & 0.95 & 0.88 \\
\texttt{L2D\_CACHE\_REFILL\_PRF} & 1017222150 & 1024578925 & 0.99 & 964186753 & 897632586 & 1.07 & 0.95 & 0.88 \\
\texttt{L2D\_CACHE\_REFILL\_DM} & 810565473 & 875772278 & 0.93 & 1010879035 & 1011919331 & 1.00 & 1.25 & 1.16 \\
\texttt{L2D\_CACHE\_WB} & 474436183 & 510418295 & 0.93 & 481165262 & 469429725 & 1.02 & 1.01 & 0.92 \\
\hline
\texttt{LD\_COMP\_WAIT} & 783957843900 & 670383891683 & 1.17 & 515460270334 & 521930831503 & 0.99 & 0.66 & 0.78 \\
\texttt{LD\_COMP\_WAIT\_EX} & 429528457728 & 323839573426 & 1.33 & 38884573461 & 37007309796 & 1.05 & 0.09 & 0.11 \\
\texttt{LD\_COMP\_WAIT\_PFP\_BUSY} & 29585337 & 14584447 & 2.03 & 101373 & 182984 & 0.55 & 0.003 & 0.01 \\
\texttt{LD\_COMP\_WAIT\_L1\_MISS} & 106374292021 & 113150011609 & 0.94 & 274005118311 & 274611290445 & 1.00 & 2.58 & 2.43 \\
\texttt{LD\_COMP\_WAIT\_L2\_MISS} & 19383905950 & 19975680958 & 0.97 & 40911779437 & 41839552962 & 0.98 & 2.11 & 2.09 \\
\texttt{EU\_COMP\_WAIT} & 625493183646 & 630338181042 & 0.99 & 340261838023 & 341315363810 & 1.00 & 0.54 & 0.54 \\
\texttt{BR\_COMP\_WAIT} & 23654302037 & 21426208561 & 1.10 & 1701837812 & 1781324275 & 0.96 & 0.07 & 0.08 \\
\texttt{BR\_MIS\_PRED} & 1941165909 & 3941304832 & 0.49 & 624100572 & 624524801 & 1.00 & 0.32 & 0.16 \\
\hline
\texttt{EFFECTIVE\_INST\_SPEC} & 1641901478240 & 1346724114727 & 1.22 & 835329678475 & 830699232211 & 1.01 & 0.51 & 0.62 \\
\texttt{LD\_SPEC} & 326350027305 & 269139241392 & 1.21 & 195567194613 & 194711154297 & 1.00 & 0.60 & 0.72 \\
\texttt{BASE\_LD\_REG\_SPEC} & 215301026770 & 158090240857 & 1.36 & 68445331297 & 67589290981 & 1.01 & 0.32 & 0.43
\end{tabular}
\end{sidewaystable*}

\begin{sidewaystable*}[]
\centering
\caption{Derived rates and their ratios for the 3-d Hydro test on 1 core. The first 3 columns show the rates for the GCC compiler with and without hp, followed by the rate ratio. The next three columns show the same, but for the Fujitsu compiler. The last 2 columns show the ratio of rates between Fujitsu : GCC compiler.}
\label{tab:app_3dHydro_Rate}
\begin{tabular}{lrrr|rrr|rr}
\textbf{Rate} & \textbf{GCC} & {\textbf{GCC}} & {\textbf{GCC}} & {\textbf{Fujitsu}} & {\textbf{Fujitsu}} & \textbf{Fujitsu} & \textbf{Fuj : GCC} & \textbf{Fuj : GCC} \\
& \textbf{hp} & \textbf{no hp} & {\textbf{hp : no hp}} & \textbf{hp} & {\textbf{no hp}} & \textbf{hp : no hp} & \textbf{hp} & \textbf{no hp} \\
\hline \hline
Cycles per instruction (CPI) & 1.31 & 1.45 & 0.90 & 1.43 & 1.45 & 0.99 & 1.10 & 1.00 \\
Branch misprediction rate & 0.001 & 0.003 & 0.40 & 0.0007 & 0.0008 & 0.99 & 0.63 & 0.26 \\
\hline
L1-ITLB miss rate & 3.62E-06 & 0.0001 & 0.03 & 4.49E-06 & 6.97E-06 & 0.64 & 1.24 & 0.07 \\
L1-DTLB miss rate & 0.0001 & 0.001 & 0.09 & 0.0006 & 0.002 & 0.31 & 4.96 & 1.39 \\
L2-ITLB miss rate & 5.13E-08 & 1.05E-07 & 0.49 & 6.43E-08 & 3.98E-07 & 0.16 & 1.25 & 3.78 \\
L2-DTLB miss rate & 1.36E-07 & 4.77E-06 & 0.03 & 2.49E-07 & 7.89E-06 & 0.03 & 1.84 & 1.65 \\
\hline
L1D cache miss rate & 0.02 & 0.02 & 0.82 & 0.03 & 0.03 & 0.99 & 1.80 & 1.49 \\
... attributable to demand access & 0.01 & 0.01 & 0.82 & 0.02 & 0.02 & 0.99 & 2.10 & 1.74 \\
... attributable to prefetch access & 0.005 & 0.006 & 0.83 & 0.005 & 0.006 & 0.98 & 1.16 & 0.99 \\
... attributable to software prefetch access & 0.0003 & 0.0004 & 0.82 & 0.0002 & 0.0002 & 0.92 & 0.51 & 0.45 \\
L2 cache miss rate & 0.001 & 0.001 & 0.79 & 0.002 & 0.002 & 1.03 & 2.12 & 1.63 \\
... attributable to demand access & 0.0005 & 0.0007 & 0.76 & 0.001 & 0.001 & 0.99 & 2.45 & 1.87 \\
... attributable to prefetch access & 0.0006 & 0.0008 & 0.81 & 0.001 & 0.001 & 1.07 & 1.86 & 1.42 \\
... attributable to software prefetch access & 0.0 & 0.0 & N/A & 5.92E-07 & 5.78E-07 & 1.03 & N/A & N/A \\
\hline
Average latency of  &  &  &  &  &  &  &  &  \\
... L1D cache miss processing & 40.82 & 40.97 & 1.00 & 42.27 & 42.16 & 1.00 & 1.04 & 1.03 \\
... L2 cache miss processing & 1.26 & 1.23 & 1.03 & 1.12 & 1.16 & 0.96 & 0.89 & 0.94 \\
Average number of outstanding misses  &  &  &  &  &  &  &  &  \\
... in L1D cache miss processing & 0.55 & 0.60 & 0.91 & 0.93 & 0.93 & 1.00 & 1.70 & 1.54 \\
... in L2 cache miss processing & 0.001 & 0.001 & 0.90 & 0.002 & 0.002 & 1.00 & 1.71 & 1.55 \\
Bidirectional effective BW between (GB/s) &  &  &  &  &  &  & &  \\
... L1D cache and L2 cache & 10.51 & 11.50 & 0.91 & 17.15 & 17.17 & 1.00 & 1.63 & 1.49 \\
... L2 cache and memory & 0.49 & 0.57 & 0.87 & 0.95 & 0.91 & 1.04 & 1.91 & 1.61 \\
\hline
\texttt{LD\_COMP\_WAIT/CPU\_CYCLES} & 0.37 & 0.34 & 1.07 & 0.43 & 0.43 & 0.99 & 1.18 & 1.27 \\
\texttt{LD\_COMP\_WAIT\_L1\_MISS/CPU\_CYCLES} & 0.05 & 0.06 & 0.86 & 0.23 & 0.23 & 1.00 & 4.62 & 3.95 \\
\texttt{LD\_COMP\_WAIT\_L2\_MISS/CPU\_CYCLES} & 0.01 & 0.01 & 0.88 & 0.03 & 0.03 & 0.98 & 3.78 & 3.41 \\
\texttt{LD\_COMP\_WAIT\_EX/CPU\_CYCLES} & 0.20 & 0.17 & 1.21 & 0.03 & 0.03 & 1.05 & 0.16 & 0.19 \\
\texttt{LD\_COMP\_WAIT\_PFP\_BUSY/CPU\_CYCLES} & 1.38E-05 & 7.46E-06 & 1.85 & 8.47E-08 & 1.52E-07 & 0.56 & 0.006 & 0.02
\end{tabular}
\end{sidewaystable*}

\end{document}